\documentclass[aps,prl,twocolumn,superscriptaddress,showpacs]{revtex4-1}
\makeatletter

\newcommand{\Rmnum}[1]{\expandafter\@slowromancap\romannumeral #1@}
\makeatother
\usepackage{graphicx}
\usepackage{latexsym}
\usepackage{amssymb}
\usepackage{amsfonts}
\usepackage{soul}
\usepackage{color}
\usepackage{amsmath}
\usepackage{natbib}
\usepackage{lineno}
\begin{document}

\title{Nanoscale assembly of superconducting vortices with scanning tunnelling microscope tip}

\author{Jun-Yi Ge}\email{Junyi.Ge@kuleuven.be}
\affiliation{INPAC--Institute for Nanoscale Physics and Chemistry,
KU Leuven, Celestijnenlaan 200D, B--3001 Leuven, Belgium}
\author {Vladimir N. Gladilin}
\affiliation{INPAC--Institute for Nanoscale Physics and Chemistry,
KU Leuven, Celestijnenlaan 200D, B--3001 Leuven, Belgium}
\affiliation{TQC--Theory of Quantum and Complex Systems,
Universiteit Antwerpen, Universiteitsplein 1, B--2610 Antwerpen,
Belgium}
\author{Jacques Tempere}
\affiliation{TQC--Theory of Quantum and Complex Systems,
Universiteit Antwerpen, Universiteitsplein 1, B--2610 Antwerpen,
Belgium}
\author{Cun Xue}
\affiliation{INPAC--Institute for Nanoscale Physics and Chemistry,
KU Leuven, Celestijnenlaan 200D, B--3001 Leuven, Belgium}
\affiliation{School of Mechanics, Civil Engineering and Architecture, Northwestern Polytechnical University, Xi'an 710071, China}
\author{Jozef T. Devreese}
\affiliation{TQC--Theory of Quantum and Complex Systems,
Universiteit Antwerpen, Universiteitsplein 1, B--2610 Antwerpen,
Belgium}
\author{Joris Van de Vondel}
\affiliation{INPAC--Institute for Nanoscale Physics and Chemistry,
KU Leuven, Celestijnenlaan 200D, B--3001 Leuven, Belgium}
\author{Youhe Zhou}
\affiliation{School of Aeronautics, Northwestern Polytechnical University, Xi'an 710071, P.R. China}
\author{Victor V. Moshchalkov}\email{Victor.Moshchalkov@kuleuven.be}
\affiliation{INPAC--Institute for Nanoscale Physics and Chemistry,
KU Leuven, Celestijnenlaan 200D, B--3001 Leuven, Belgium}

\date{\today}

\begin{abstract}

\end{abstract}

\maketitle

\textbf{Vortices play a crucial role in determining the
properties of superconductors as well as their
applications. Therefore, characterization and manipulation of
vortices, especially at the single vortex level, is of great
importance. Among many techniques to study single vortices, scanning
tunneling microscopy (STM) stands out as a powerful tool, due to its
ability to detect the local electronic states and high
spatial resolution. However, local control of superconductivity as well as the manipulation of individual
vortices with the STM tip is still lacking. Here we report a new
function of the STM, namely to control the local pinning in a
superconductor through the heating effect. Such effect allows us to
quench the superconducting state at nanoscale, and leads to the
growth of vortex-clusters whose size can be controlled by the bias
voltage. We also demonstrate the use of an STM tip to assemble
single quantum vortices into desired nanoscale configurations.}

Pair condensation of charge carriers from the normal to the
superconducting state is associated with a gain of the free energy.
However, in the presence of a magnetic field, this gain is decreased
due to the expulsion of the external field by the superconductor.
For type-II superconductors, this competition results in the
penetration of quantized vortices which are characterized by two
length scales: the coherence length $\xi$, equivalent to the size of
the vortex core, and the magnetic penetration depth $\lambda$, the
decay length for the supercurrents encircling a vortex core
\cite{Blatter}. Since the formation of the vortex core requires
breaking of Cooper pairs, it is energetically favorable for a vortex
to be placed in an area where superconductivity is already locally
suppressed. Such positions are called pinning centers and have
various forms such as atom vacancies \cite{Liang}, variations of
sample composition or/and thickness \cite{Bezryadin} and
artificially patterned antidots \cite{Moshchalkov1998,Silva}. The
manipulation and control of pinning configuration and its
interaction with vortex are fundamental for creating new functional
devices with superconductors \cite{Auslaender,MO,Bending-SHPM}.

The scanning tunnelling microscope (STM) is one of the most powerful
techniques to image nanoscale topography and characterize the local
physical properties \cite{Suderow,Roditchev,Shan,Sandhu}. It has
been widely used in the study of condensed matter systems such as
insulators \cite{Wong}, semiconductors \cite{Butler} and
superconductors \cite{Zhao}. Also the capability of precise
positioning  makes it a powerful tool to design various artificial
nanostructures at atomic level \cite{Eigler,Eigler-2,Kalff,Tanaka}.
In the study of superconductivity, the STM can probe the local
electronic density of states and thus is used to map the spatial
distribution of superconducting gap and the vortex core state. The
STM is the most suitable technique to directly image vortices with
atomic resolution, especially at high magnetic fields. A series of
intricate phenomena related to vortex states have been revealed by
the STM, such as the vortex lattice melting process
\cite{Guillamon}, the anisotropy of the Fermi surface distribution
of the superconducting gap \cite{Song}, order-disorder transition
\cite{Guillanmon2}. However, despite the importance of technical and
scientific applications by utilizing vortices to design fluxonic
devices \cite{Golod,Miyahara}, local control of superconductivity
and the precise manipulation of vortices with the STM is still
lacking.

\begin{figure*}[!t]
\centering
\includegraphics*[width=1\linewidth,angle=0]{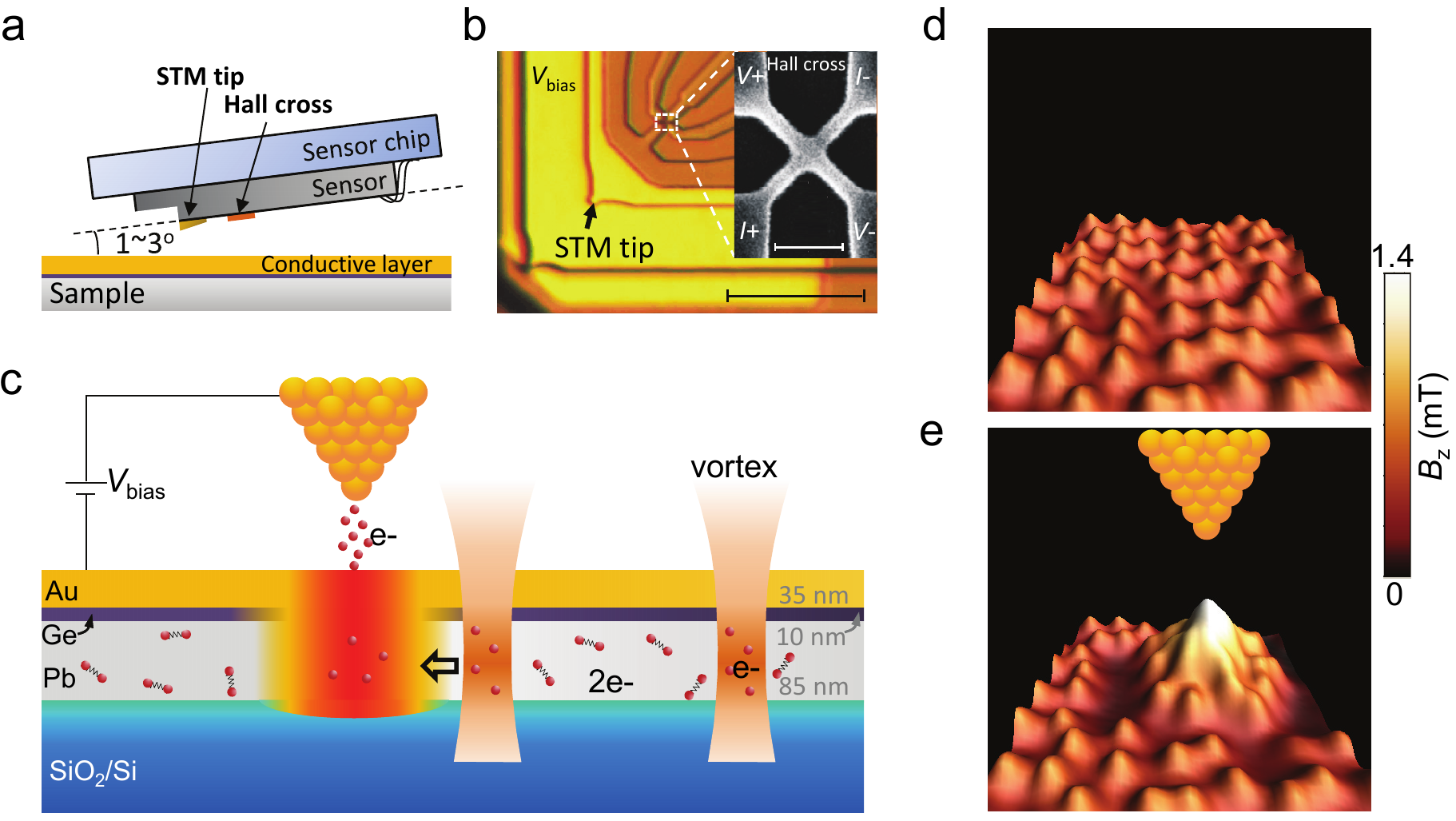}
\caption{\textbf{Introduction to the operation of the heating effect
generated with a scanning tunnelling microscope tip.} (\textbf{a})
Schematic view of the scanning Hall probe microscope (SHPM). A scanning tunneling microscope (STM) tip is
assembled together with a Hall cross to make a sensor, which is
aligned at a small angle (1 to 3 degrees) to the sample surface.
(\textbf{b}) An optical image of the Hall sensor. An SEM image of
the Hall cross is shown in the inset. The longer (shorter) scale bar corresponds to 20 (1) $\mu$m. (\textbf{c}) Schematic
representation of the local heating effect by using the STM tip. The
area close to the tip is heated up by the tunnelling current while
the insulating Ge layer and the superconducting Pb layer are also
warmed up due to thermal transfer. Superconductivity is suppressed
in a localized region where it is energetically favorable to place
vortices. (\textbf{d}) SHPM image of a vortex lattice observed after
field cooling at $H=3.74\times10^2$ Am$^{-1}$ from above $T_\textrm{c}$ down to
$T=4.2$ K. (\textbf{e}) SHPM image after 5 seconds of tunnelling at
bias voltage of 0.5 V and tunnelling current of 0.5 nA, and then
lifting up the tip for Hall probe imaging. A vortex cluster forms at
the tip position due to the local quench of the hot spot.}
\label{fig1}
\end{figure*}

Here, we report a method of controlled quenching of a hot spot in a
superconducting film by using the local heating effect of the
tunnelling junction on vortex states, which is especially
interesting at the nanoscale. This method utilizes the heating
effect of the tunneling junction with the power which can be well
controlled by tuning the bias voltage. We are also able to use this
technique to precisely manipulate single quantum vortices. Our
results extend the pioneering work of Eigler and Schweizer \cite{Eigler} from
manipulating single atoms to manipulating single quantum vortices.

$\\$ \textbf{Results}$\\$ \textbf{Sample design and vortex
assembling}. The experiments were performed by imaging vortices
using scanning Hall probe microscopy (SHPM) which combines the
non-invasive Hall imaging technique and an STM (Figs.~1a, b). The
STM tip, typically used only to provide feedback to bring the Hall
sensor in close proximity to the sample surface \cite{Bending}, is
here used as a local heating source. In the presence of tunnelling
current, a pinning potential is created at the position of the tip
due to the local suppression of the superconductivity. As a result,
nearby vortices will be attracted to this region. However, once the
tunnelling disappears by retracting the tip to a certain distance
(200~nm), the heating effect is stopped and the
accumulated vortices will finally relax into a triangular vortex
lattice due to the repulsive interaction among them. To avoid this
and to retain the topological defects after quenching, a sample with
sufficiently strong pinning centers and weak vortex-vortex
interactions is required.

The sample schematically shown in Fig.~1c is a multilayered structure of
Pb/Ge/Au deposited on a SiO$_2$/Si substrate (see Methods). The superconducting Pb film, which is characterized by relatively weak vortex-vortex interactions and a high density of quasi-homogeneously distributed pinning centers, is an ideal candidate to study the local heating effect. The pinning centers are created naturally during the sample preparation and can be considered as locations with a reduced electron mean free path \cite{Ge}. The vortex-vortex interaction can be weakened by choosing a superconducting film with relatively small Ginzburg-Landau parameter $\kappa=\lambda/\xi$. Therefore, a Pb film with thickness $d=85$~nm, $\lambda(0)=94$~nm and $\xi(0)=52$~nm is used (Supplementary Figure 1, 2 and 3, and Supplementary Notes 1 and 2). In this case, once vortices are forced together, the attraction between pinning centers and vortices overcomes the repulsive interaction between vortices, so that a specific vortex configuration, formed in the course of fast hot-spot quenching, can be preserved.

\begin{figure*}[!t]
\centering
\includegraphics*[width=1\linewidth,angle=0]{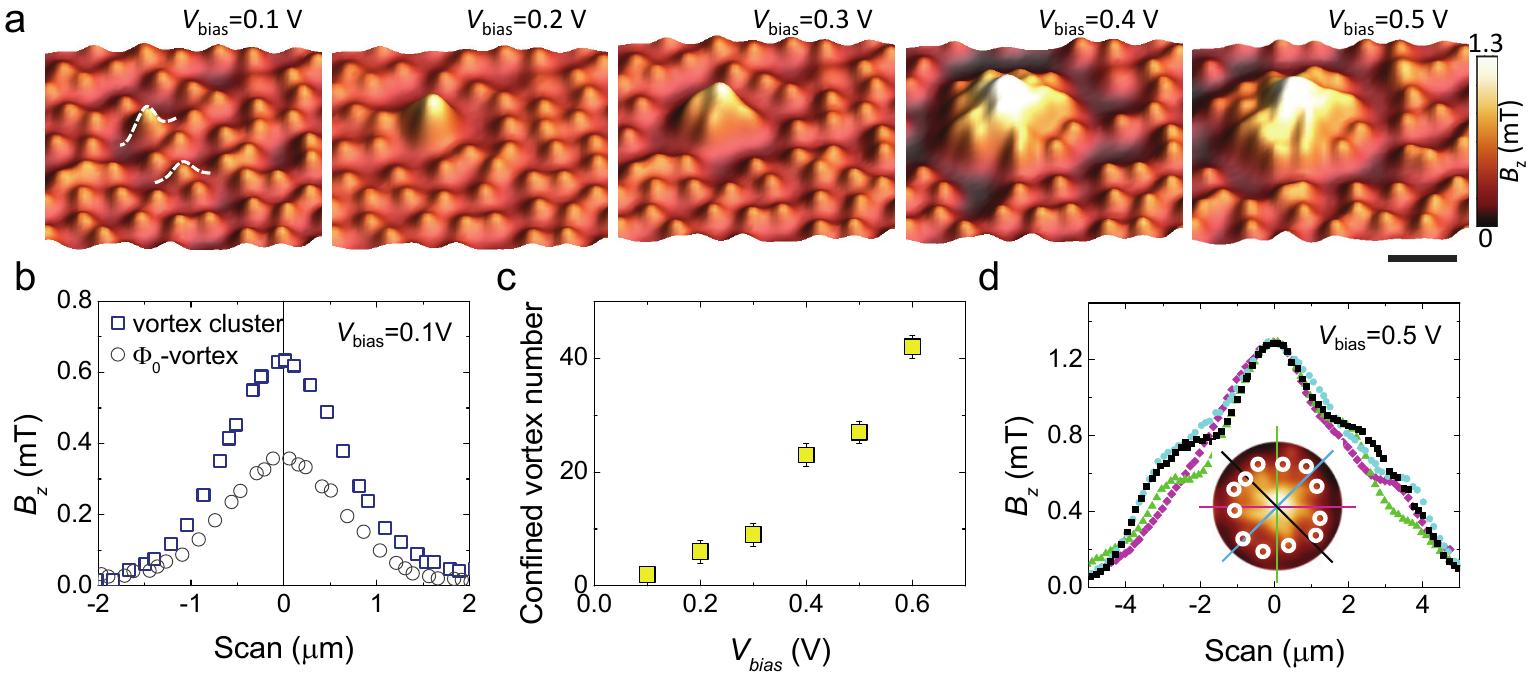}
\caption{\textbf{Vortex clustering as a function of bias voltage.}
(\textbf{a}) Scanning Hall probe microscopy (SHPM) images observed after tunneling pulses with various bias voltages $V_{\textrm{bias}}$. The size of the vortex cluster
increases with bias voltage. The minimum cluster that could be
observed contains two flux quanta at $V_\textrm{bias}=0.1$ V. The scale bar equals 4 $\mu$m. (\textbf{b})
Field profiles across the center of an individual vortex (circles)
and the smallest vortex cluster (squares), as indicated by the
dashed lines in (\textbf{a}), observed at $V_\textrm{bias}=0.1$~V.
(\textbf{c}) Number of vortices in the cluster as a function of the
bias voltage (Supplementary Figure 4 and Supplementary Note 3). The error bar corresponds to the number of vortices cut by the edges of the scanned SHPM images. (\textbf{d}) Magnetic field profiles along different
directions across the center of a vortex cluster observed at
$V_\textrm{bias}=0.5$~V. All the profiles are similar to each other,
suggesting that the vortex distribution in the cluster is approximately
axially symmetric. The white circles mark the position of individual
vortices at the periphery of the cluster (see Supplementary Figure 5).
} \label{fig2}
\end{figure*}

Figure~1d presents the vortex distribution SHPM image after
performing field-cooling (FC) from above $T_\textrm{c}$ to 4.2 K. A
disordered vortex lattice, homogeneously distributed in the scanned
area, is observed. After applying a pulse of tunnelling voltage,
followed by the Hall probe imaging, a vortex cluster is observed at
the STM tip position. Since there is no multi-quantum vortex
observed in our sample, we assume that the vortex cluster is
composed of single quantum vortices (see Supplementary Note 4).
Between the vortex cluster and the rest of individual vortices a
vortex-free area is formed. Note that the positions of the vortices
outside the cluster remain unchanged. This confirms that the impact
of heating is local. The vortex cluster retains the same structure
even after a few hours relaxation, suggesting that it is a
relatively stable configuration. Similar results have been observed
in three different samples. It has been reported that the STM
tunnelling junction can reshape the topography of a material,
generating defects at the tip position \cite{Baek}. This possibility
is ruled out since a disordered lattice is again observed in the
same region after a subsequent FC procedure. The possibility of
mechanically induced vortex attraction \cite{Nanoletter} is also
ruled out since, in our case, there is no direct contact between the
STM tip and the sample.

\begin{figure}[!t]
\centering
\includegraphics*[width=1\linewidth,angle=0]{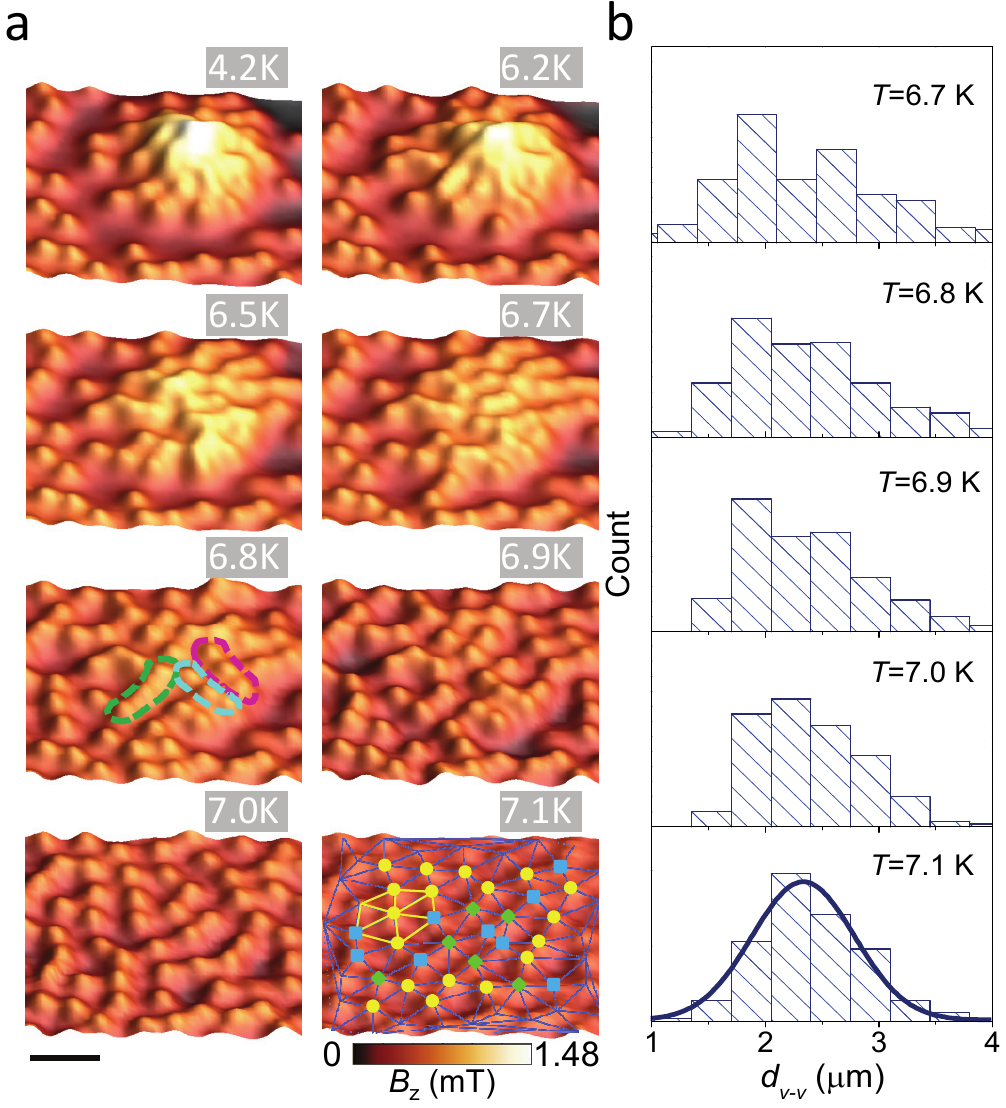}
\caption{\textbf{Disintegration of vortex cluster with increasing
temperature.} (\textbf{a}) Scanning Hall probe microscopy (SHPM) images observed after  forming a
vortex cluster at 4.2 K and then increasing the temperature
gradually. The vortex cluster remains compact until $T=6.2$ K. With
further increasing temperature, the vortex cluster expands due to
the weakening of pinning and finally a homogeneous vortex lattice is
observed at $T=7.1$ K. The Delaunay triangulation of the vortex
pattern at $T=7.1$ K is shown, where the vertices of each triangle
represent the locations of the vortices. The squares, circles and
diamonds represent the vortices with five-, six- and seven-fold
symmetries. The scale bar equals 4 $\mu$m. (\textbf{b}) Histograms of the vortex nearest neighbor
distance obtained from the images in (\textbf{a}). The distribution
at 7.1 K can be well simulated using the Gaussian curve.
}\label{fig3}
\end{figure}

$\\$ \textbf{Controlling the vortex cluster size}. The size of the
vortex clusters can be well controlled. Figure~2a shows a series of
images measured after applying the pulse of tunnelling current at
the same position under various bias voltages. It is clear that the
size of the vortex cluster increases with increasing bias voltage.
The smallest vortex cluster is observed at $V_\textrm{bias}=0.1$~V, below
which no vortex clustering can be seen. Figure~2b compares the
magnetic field profiles through the centers of the vortex cluster
observed at $V_\textrm{bias}=0.1$~V and a single-quantum vortex. It is seen
that the magnetic field at the center of the cluster is doubled as
compared to a single-quantum vortex, suggesting that it contains two
flux quanta (Supplementary Note 4). The number of trapped
vortices as a function of the bias voltage is shown in Fig.~2c. The
power dissipated in the sample is $P\propto V_\textrm{bias}I$, with $I$
being the tunnelling current which is kept constant in our
experiments. Correspondingly, the size of the region, where
superconductivity is suppressed by local heating, is expected to
increase with bias voltage. This accounts for the observed
dependence in Fig.~2c.

One clear feature of the observed vortex clusters is that, instead
of forming a disordered Abrikosov vortex lattice, clustered vortices
are arranged into concentric ring-like structures. As demonstrated
in Fig.~2d, the field profiles through the center of a vortex
cluster are nearly symmetric with a peak at the cluster center. Such
kind of ring-shaped vortex clusters, predicted by Shapiro et al
\cite{Shapiro}, might be attributed to the Kibble-Zurek (KZ) symmetry-breaking
phase transition \cite{Kibble,Zurek}. When the tunnelling current is
applied, the local area is heated up above $T_\textrm{c}$, leading to the
formation of a normal domain with homogeneous distribution of
magnetic flux in this domain. After the heating is stopped by
retracting the STM tip, the recovery of the superconductivity starts
when the temperature front deviates from the order parameter front.
The temperature front accelerates while the order parameter front,
which presses the magnetic flux  confined inside the shrinking
normal region, decelerates, leading to an unstable normal domain
ring area with $T<T_\textrm{c}$ and $\psi= 0$. As a result, vortices nucleate
in this normal domain ring at the perimeter of the heated area and
the hot spot looses part of its magnetic flux. The order parameter
front then accelerates briefly, before the above-described dynamics
repeat \cite{Shapiro,Groger}. With further decreasing temperature,
natural pinning in the heated area becomes evident again, and the
ring shaped vortex cluster is preserved even when the temperature of
the whole area is far below $T_\textrm{c}$.  As far as we know, the
ring-shaped vortex clusters have never been observed experimentally
with single vortex resolution. The possible relation of our
experiment to the KZ mechanism is further supported by the
theoretical simulations, where vortex-antivortex pairs, as
topological defects, appear from quenching the hot spot
(Supplementary Figure 6, 7 and 8, and Supplementary Note 5). However, due to the short annihilation
time, the stabilization of KZ vortices and antivortices have not
been observed experimentally. Further experimental evidence is
needed to clarify the mechanism, such as the relation of KZ vortices
with quenching rate.

As mentioned above, the vortex clusters are preserved due to the
naturally formed pinning centers. The pinning strength is weakened
when approaching $T_\textrm{c}$ \cite{Fietz}. Therefore, we expect a
competition between pinning and vortex lattice elasticity with
varying temperature. This also provides an opportunity to study the
relaxation of local non-equilibrium configuration in a macroscopic system.
Figure~3a displays the vortex cluster evolution with increasing
temperature. When the vortex cluster is created at 4.2 K, it
preserves the same geometry up to $T=6.2$~K (see Supplementary
Figure 9), above which the pinning strength is considerably weakened
as compared to the vortex-vortex interaction and the vortex cluster
starts to disintegrate. We notice that the disintegration process
starts from the interior of the vortex cluster, where vortex density
is high, so that the vortex-vortex interaction is strong. At
intermediate temperatures (e.g., $T=6.8$~K), vortices tend to form
chains as highlighted by the dashed lines in Fig.~3a. Such vortex
chains might be associated with the one-dimensional
temperature-induced vortex motion as observed in the vortex lattice
melting process \cite{Guillamon}. Only at high enough temperature
(e.g., $T=7.1$~K), vortex repulsion strongly dominates and a
homogeneous distribution of vortices, corresponding to a weakly
distorted Abrikosov lattice, is formed. Additional information about
the vortex cluster disintegration process can be obtained by
considering the nearest neighbor distance (NND) of vortices. As
presented in Fig.~3b, the NND shows a broad distribution at low
temperatures. At $T=7.1$ K, it can be well fitted by the Gaussian
distribution (solid line). The observed peak position at
$d_\textrm{v-v}=2.3~\mu$m is consistent with the value expected for an
Abrikosov triangular vortex lattice,
$d_\textrm{v-v}=(2\Phi_0/\sqrt{3}H_0)^{1/2}=2.27$~$\mu$m.

$\\$ \textbf{Theoretical simulations}. The proposed scenario of
vortex clustering is supported by the results of numerical
simulations, based on the time-dependent Ginzburg-Landau (TDGL)
formalism and the local heating model (Supplementary Figure 10 and Supplementary Note 6). The
deviation of the local temperature $T$ from its equilibrium value in
the absence of tunnelling current, $T_0$,  is represented as
$T-T_0=\alpha f(x,y)$ (see Fig. 4a), where the coefficient $\alpha$
is proportional to the power dissipated by the tunnelling current,
while the spatial temperature distribution $f(x,y)$ is determined by
material and geometric parameters of the sample. Using this
temperature distribution to model the effect of a tunnelling current
pulse, the TDGL calculations are performed for a $10 \times
10$~$\mu$m$^2$ superconductor film with random pinning centers (Fig.
4b). Based on the used local heating model and the material
parameters, the quenching time $\tau_{\rm Q}^{-1}=-T^{-1}\partial
T/\partial t |_{T=T_\textrm{c}}$, corresponding to the thermal relaxation of
the hot spot after switching off the tunneling current, is estimated
to lie within the range of 1 to 10 ps for the initial radii of the
normal domain in the hot spot from 1 to 3 $\mu$m. These
quenching-time values are comparable to the Ginzburg-Landau
characteristic time in our sample, $t_\textrm{GL}(T_0) = 1$ ps.  The
formation of vortex clusters, resulting from the simulated quenching
process, and their stabilization due to vortex pinning typically
occurs on the time scale $\sim40$ to 100 ps. The vortex cluster
formation time, compared with the short quenching time, clearly
places the quenching process in the KZ regime.

\begin{figure}[!t]
\centering
\includegraphics*[width=1\linewidth,angle=0]{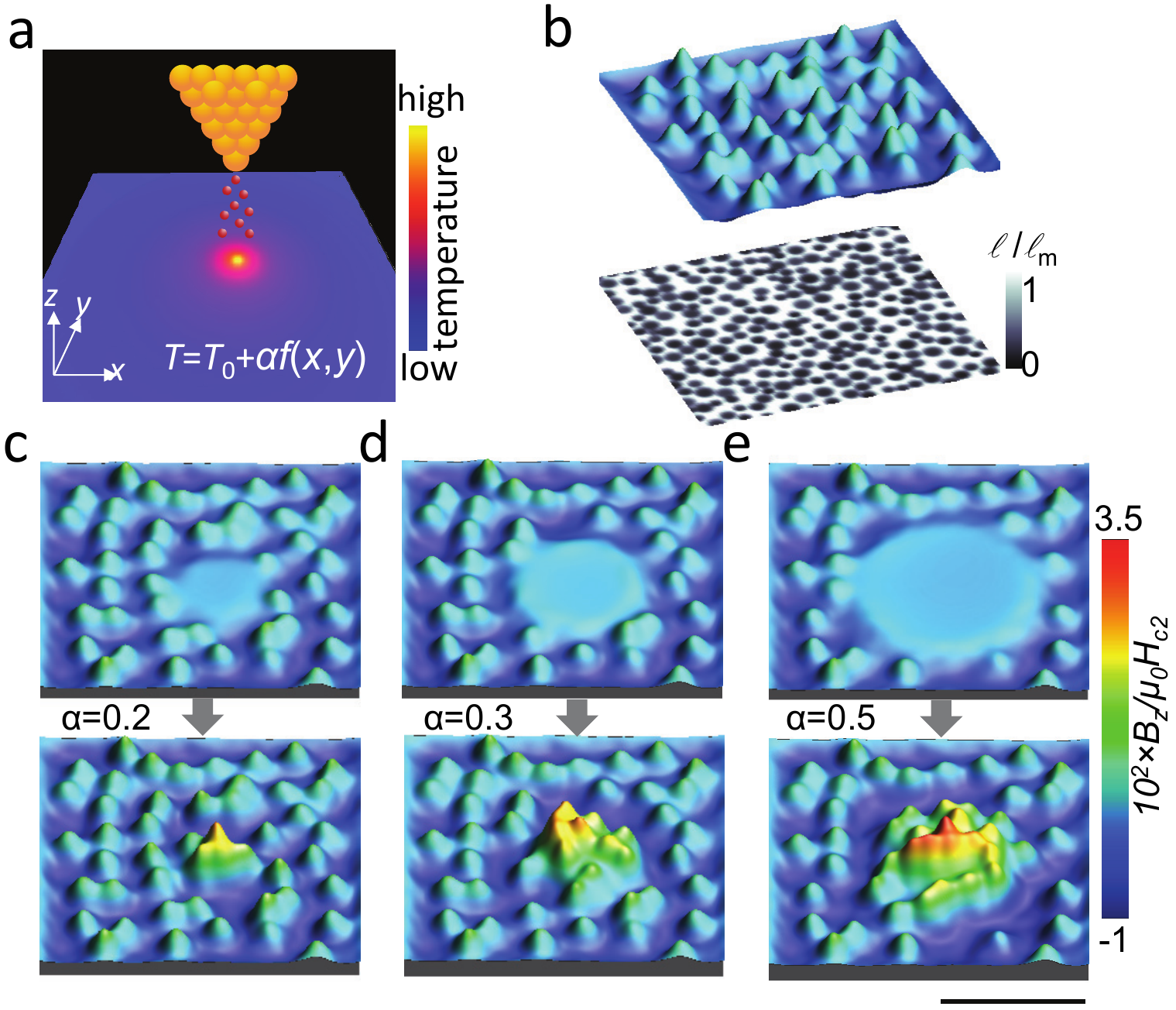}
\caption{\textbf{Simulations.} (\textbf{a}) Temperature distribution
at the scanning tunneling microscope tip position. (\textbf{b})
Vortex distribution (upper panel) after field cooling down to the
temperature $T_0=0.58T_\textrm{c}$ at magnetic field $H_0=0.03H_\textrm{c2}(T_0)$.
The lower panel shows the distribution of the electron mean free
path, which models a set of quasi-random pinning centers. The
magnetic field distributions, which are formed in the course of a
tunnelling current pulse (upper panel) and after switching off the
pulse (lower panel) are shown for the parameter $\alpha=0.2$
(\textbf{c}), 0.3 (\textbf{d}) and 0.5$T_\textrm{c}$ (\textbf{e}). When the
tunnelling current is on, local superconductivity is fully
suppressed. The scale bar equals 4 $\mu$m.} \label{fig2}
\end{figure}

The simulation results are displayed in Figs. 4c to 4e for different
values of the coefficient $\alpha$. The local heating, caused by
tunnelling current, leads to the formation of a normal-state region
in the vicinity of the STM tip. The normal region can be considered
as a big pinning site, which is capable to accommodate a relatively
large magnetic flux. This pinning site attracts and traps vortices
from the surrounding superconducting area, where the temperature is
elevated and the mobility of vortices is increased due to their
weaker interaction with pinning centers. The magnetic flux trapped
within the normal region increases with increasing $\alpha$. When
the tunneling current is switched off and the local temperature
relaxes to $T_0$, the trapped flux transforms into a compact vortex
cluster, which is stabilized by the pinning centers. Similarly to
the experiment, the calculated vortex pattern demonstrates the
appearance of vortex-depleted regions around the vortex cluster,
which become more evident when increasing the intensity of local
heating and the cluster size.

\begin{figure}[!t]
\centering
\includegraphics*[width=1\linewidth,angle=0]{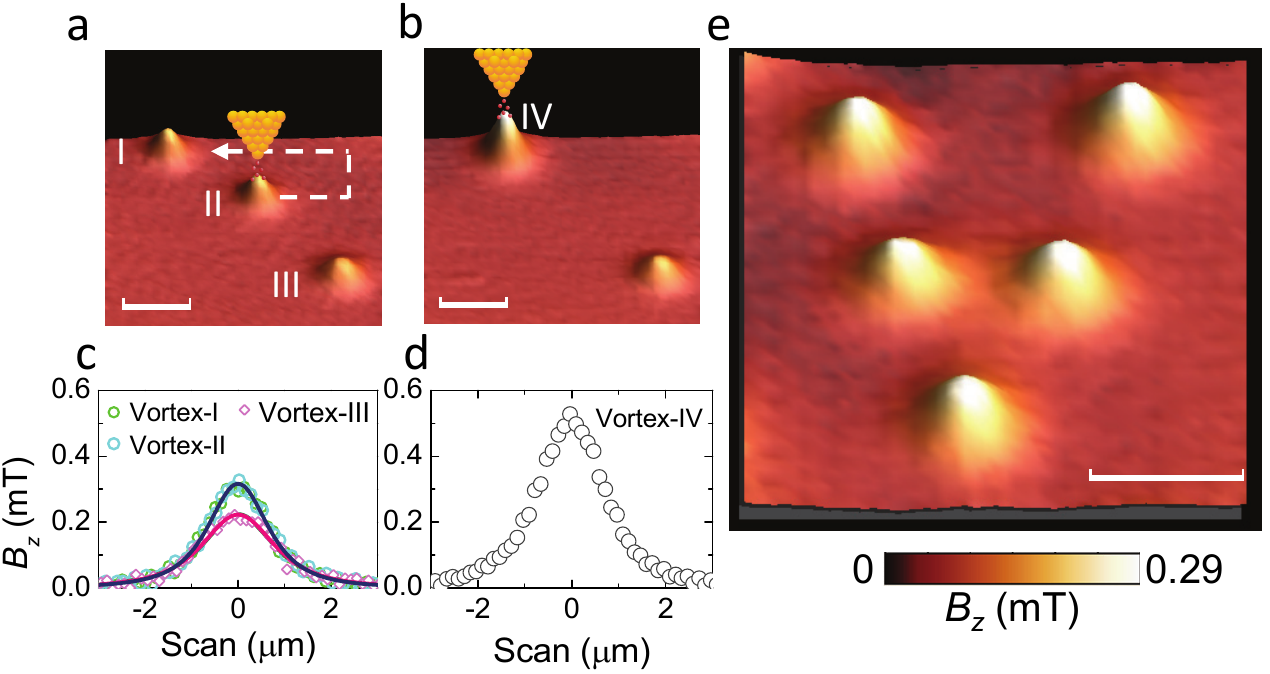}
\caption{\textbf{Manipulation of individual vortices with the scanning tunneling microscopy (STM)
tip.} (\textbf{a}) SHPM image observed after field cooling from
above $T_\textrm{c}$ to $T=4.2$ K at $H_0=4.3$ Am$^{-1}$. Three single flux quantum
vortices are observed at naturally formed pinning centers.
(\textbf{b}) Image observed after moving  vortex II with the STM tip
along the arrow in (\textbf{a}). Vortices I and II merged to form a
vortex cluster IV.  (\textbf{c, d}), Magnetic field profiles through
the centers of vortices I, II, III and vortex cluster VI. The solid
lines are a fit with the monopole model. (\textbf{e}), A pattern
V short for vortex is arranged by using the STM tip.The scale bars equal 4$\mu$m for all the images.}
\label{fig2}
\end{figure}

$\\$ \textbf{Manipulation of single-quantum vortices}. The obtained
results suggest that local heating with an STM tip can provide an
efficient tool to arrange vortices in superconductors. Finally, we
demonstrate the ability of using the STM tip to manipulate
individual vortices (see also Supplementary Movie). This is an
important extension of the pioneering work of Eigler \cite{Eigler}
from manipulating single atoms to manipulating single quantum
vortices. To move a vortex, first, a tunneling junction is
established by approaching the STM tip to the surface of the sample
at 1~$\mu$m away from the vortex center. A hot spot is generated
and the vortex is attracted to the location of the hot spot. Then
the STM tip is retracted and moved to the next position. By
repeating the above process, we are able to drag the vortex to any
position in the superconductor. As shown in Fig.~5a, a single
quantum vortex (vortex II) can be moved along the path indicated by
the dashed arrow to merge with vortex I, forming a vortex cluster
(IV in Fig.~5b). We are also able to drag the vortex cluster as one
object to a new position and then detach the two vortices by
increasing the temperature (see Supplementary Figure 11). By using the
local heating method demonstrated above, a pattern V, short for
vortex, is arranged from $\Phi_0$-vortices as shown in Fig.~5e.
Compared with other techniques, such as magnetic force microscopy
\cite{Auslaender}, that are used to manipulate individual vortices,
the big advantage of our technique is that no magnetic field or
transport current is needed.

$\\$ \textbf{Discussion}$\\$ In conclusion, we have shown that a
controlled heating effect generated by the tunnelling current can be
used to locally suppress superconductivity, with this area playing
the role of a pinning center.
We are able to tune to power of the nano heater simply by changing the bias voltage.
Moreover, the tunnelling pulse applied through the STM tip enables
dragging and manipulation of individual quantized vortices which may
be used for the development of new fluxon-based devices. Our results
demonstrate the utility of the local heating effect for
characterizing and manipulating vortices in a superconductor. This new approach provides a lot of information which is of great interest for studying other systems,
such as local and non-local phase transitions in superfluid,
cosmology and many-particle systems.

$\\$

$\\$
\textbf{Methods}\\
\textbf{Sample and experimental setup}. The trilayer of Au/Ge/Pb was
prepared in two steps. First, to ensure a relatively small $\kappa$
of the sample, a 85 nm thick Pb film was deposited on a SiO$_2$/Si
substrate using an ultra high vacuum ($3\times10^{-8}$ Torr)
electron beam evaporator calibrated with a quartz monitor. The
substrate was cooled to 77~K by liquid nitrogen to ensure the
homogeneous growth of Pb. On top of Pb, a 10 nm thick Ge layer is
deposited to protect the sample surface from oxidation and also to
avoid any proximity effect between Pb and the subsequently deposited
conductive layer \cite{Kim}. Second, the sample was transferred into
a sputtering machine, and a 35~nm thick Au layer was deposited to
cover the whole sample, playing the role of a conducting layer for
the tunnelling current of the STM tip. The thickness of Au yields a
smooth surface with a roughness less than 0.2 nm. The critical
temperature $T_\textrm{c}=7.25$~K is determined using local ac
susceptibility measurements. The pulse tunneling was applied by
controlling the bias voltage between the STM tip (Au) of a
commercial Hall probe as shown in Fig.~1b and the sample, which is
ground. In our experiments, the tunneling current is kept constant
($\sim 0.5$ nA) through a PID protocol control. The magnetic field
distribution is recorded using the scanning Hall probe microscope
from Nanomagnetics in a lift-off mode (Supplementary Figure 12 and Supplementary Note 7).
First, the Hall sensor approaches the sample under the control of a
piezo until the tunneling current is established. Then the whole
sensor is retracted by 200 nm and the magnetic field distribution is
mapped. In all the measurements, the magnetic field is applied
perpendicularly to the sample surface. WSxM software is used to
process all SHPM images \cite{wsxm}.

$\\$ \textbf{TDGL simulation}. For a superconductor with pinning
centers, which originate from a local reduction of the mean free
path $\ell$, the TDGL equation for the order parameter $\psi$,
normalized to 1, can be written in the form~\cite{Ge}
\begin{eqnarray}
\left(\frac{\partial }{\partial t }+i\varphi\right)\psi
&&=\frac{\ell}{\ell_\textrm{m}}\left( \nabla -i \mathbf{A} \right)
^{2}\psi
 \nonumber \\ &&+2\psi \left(\frac{1-T^2/T_\textrm{c}^2}{1-T_0^2/T_\textrm{c}^2}-\frac{\ell_\textrm{m}}{\ell}|\psi|^{2}\right),
\label{GLgen}
\end{eqnarray}
where $\varphi$ and $\mathbf{A}$ are the scalar and vector
potentials, respectively, and $l_\textrm{m}$ is the mean free path
value outside the pinning centers. The relevant quantities are made
dimensionless by expressing lengths in units of $\sqrt{2}\xi(T_0)$,
time in units of $\pi\hbar/[4k_\textrm{B}(T_\textrm{c}-T_0)] =2
t_\textrm{GL}(T_0)$, magnetic field in units of $\Phi_0/[4\pi
\xi^2(T_0)]=\mu_0 H_{\textrm{c2}}(T_0)/2$, and scalar potential in
units of $2k_\textrm{B}(T_\textrm{c}-T_0)/(\pi e)$. Here, $\mu_0$ is
the vacuum permeability, $t_{\textrm{GL}}$ is the Ginzburg-Landau
time, $H_{\textrm{c2}}$ is the second critical field and $\xi$ is
the coherence length at $l=l_\textrm{m}$.

The vector potential ${\bf A}$, for which we choose the gauge
$\nabla \cdot {\bf A}=0$, can be represented as ${\bf A }={\bf A }_0
+{\bf A }_1$. ${\bf A }_0$ denotes the vector potential
corresponding to the externally applied magnetic field ${\bf H}_0$,
while
 ${\bf A }_1$ describes the magnetic field ${\bf H}_1$ induced
by the currents ${\bf j}$, which flow in the superconductor:
\begin{eqnarray}
{\bf A }_{1} ({\bf r})=\frac{1}{2\pi\kappa^2}\int  d^3 r^\prime
\frac{{\bf j}({\bf r}^\prime)}{|{\bf r}-{\bf r}^\prime |}.
\label{a1}
\end{eqnarray}
Here, $\kappa=\lambda(T_0)/\xi(T_0)$ is the Ginzburg-Landau
parameter and $\lambda$ is the penetration depth  at
$\ell=\ell_\textrm{m}$. The current density is expressed in units of
$\Phi_0/[2\sqrt{2}\pi \mu_0 \lambda(T_0)^2 \xi(T_0)]=3\sqrt{3}
/(2\sqrt{2})j_\textrm{c}(T_0)$ with $j_\textrm{c}$, the critical
(depairing) current density of a thin wire or film. Integration in
Eq.~(\ref{a1}) is performed over the volume of the superconductor.

In general, the total current density contains both the
superconducting and normal components: ${\bf j}={\bf
j}_\textrm{s}+{\bf j}_\textrm{n}$ with
\begin{eqnarray}
{\bf j}_\textrm{s}={\rm Im}\left(\psi^*\nabla\psi \right)- {\bf
A}|\psi|^2, \label{js}
\end{eqnarray}
\begin{eqnarray}
{\bf j}_\textrm{n}=-\frac{\sigma}{2}\left(\nabla\varphi
+\frac{\partial \bf{A}}{\partial t }\right),  \label{jn}
\end{eqnarray}
where $\sigma$ is the normal-state conductivity, which is taken as
$\sigma=1/12$ in our units~\cite{kato91}. The distribution of the
scalar potential $\varphi$ is determined from the condition
\begin{eqnarray} \nabla \cdot {\bf j}=0,  \label{div}
\end{eqnarray}
which reflects the continuity of currents in the superconductor.
Both the $j_\textrm{n}$ and $\varphi$ vanish when approaching
(meta)stable states, which are of our main interest here.

Assuming that the thickness of the superconductor layer is
sufficiently small, variations of the order parameter magnitude
across the sample as well as currents in this direction can be
neglected and Eq.~(\ref{GLgen}) becomes effectively two-dimensional.
This equation together with Eqs.~(\ref{a1}) and (\ref{div}) are
solved self-consistently following the numerical approach described
in Ref.~\cite{silhanek2011local}.

$\\$ \textbf{Data availability}. All relevant data are available from the corresponding author.

$\\$
\textbf{Acknowledgements}\\
We acknowledge the support from the Methusalem funding by the
Flemish government, the Flemish Science Foundation (FWO) and the
MP1201 COST action. J.T. also acknowledges support from the Research
Council of Antwerp University (BOF). Y.Z. and C.X. acknowledge the
National Natural Science Foundation of China (No. 11421062), the
National Key Project of Magneto-Constrained Fusion Energy
Development Program (No. 2013GB110002) and the CSC program. $\\$
\textbf{Author contributions}\\
J.Y.G. and C.X. made the sample. J.Y.G. performed the SHPM measurements.
V.G., J.T. and J.T.D. did the simulations. J.Y.G., V.G. and V.V.M.
wrote the manuscript. All authors contributed to the discussion and
analysis. V.V.M. coordinated the whole work. $\\$
\textbf{Competing financial interests}\\
The authors declare no competing financial interests.

\end{document}